\documentclass[prd,draft,preprint,showpacs,groupedaddress]{revtex4-1}
\usepackage{amsmath}
\usepackage{amsfonts}
\usepackage{amssymb}
\usepackage{geometry}
\usepackage{natbib}
\usepackage[english]{babel}

\begin{document}

  \title{Discussion of a possible corrected black hole entropy}

  \author{Miao He} \author{Ziliang Wang} \author{Chao Fang}
  \author{Daoquan Sun} \author{Jianbo Deng}\email[Jianbo Deng: ]{dengjb@lzu.edu.cn}
  
  \affiliation{Institute of Theoretical Physics, LanZhou University,
    Lanzhou 730000, P. R. China}

  \date{\today}

  \begin{abstract}
  Einstein's equation could be interpreted as the first law of thermodynamic near the 
  spherically symmetric horizon. Through recalling the Einstein gravity with a more general
  static spherical symmetric metric, we find that the entropy would have a correction in 
  Einstein 
  gravity. By using this method, we investigate the Eddington-inspired Born-Infeld (EiBI) 
  gravity. Without matter field, we can also derive the first law in EiBI gravity. With an 
  electromagnetic field, as the field equations have a more general spherically symmetric 
  solution in EiBI gravity, we find that correction of the entropy could be generalized to EiBI
  gravity. Furthermore, we point out that the Einstein gravity and EiBI gravity might be 
  equivalent on the event horizon. At last, under EiBI gravity with the electromagnetic field, a 
  specific corrected entropy of black hole is given.       
  \end{abstract}

  

  \maketitle

  \section{INTRODUCTION}
  Black hole thermodynamics has been proposed for many years since the entropy and 
  temperature were found by Bekenstein and Hawking~\cite{1,2}, even got many interesting 
  results, like the four laws of black hole thermodynamics. It established the connection 
  between the gravity and thermodynamics. 
  \par
  The entropy is assumed to be proportional to its 
  horizon area~\cite{1}, and it is well-known that the so-called area formula of black hole 
  entropy holds only in Einstein gravity. However, when some higher order curvature terms 
  appear
  in some gravity theory, the area formula has to be modified~\cite{15}. 
  A logarithmic term often occurs in the 
  correction like the black hole entropy in loop quantum gravity (quantum geometry)~\cite
  {16,17,18} and thermal equilibrium fluctuation~\cite{19,20}. The 
  correction of entropy has been studied in Gauss-Bonnet gravity~\cite{22}, Lovelock gravity~
  \cite{23} and $f(R)$ gravity~\cite{21}. In the apparent horizon of FRW universe, the entropy 
  also has a correction~\cite{14}. 
  \par
  The Einstein's equation can be derived from the thermodynamics~\cite{3}, on the other side, 
  the thermodynamic route to the gravity field equation, which could get the first law of 
  thermodynamic in Einstein gravity, was proposed by T. Padmanabhan~\cite{4,5,6}. It 
  indicated a 
  generic connection between thermodynamics of horizons and gravity, although it's not yet 
  understood at a deeper level~\cite{13}. This technique has been used in Gauss-Bonnet gravity 
  and Lanczos-Lovelock gravity~\cite{5}, the corrected entropy is the same in ~\cite{26,27},  
  respectively.
  \par
  The EiBI gravity was inspired by Banados and Ferreira~\cite{8}. It is completely equivalent to 
  the Einstein gravity in vacuum, but in the presence of matter, it would show many interesting 
  results, like an alternative theory of Big Bang singularity in early universe~\cite{11} and 
  the mass inflation in EiBI black holes~\cite{9,10}. However, there are few investigations for
  the thermodynamic properties of black hole in EiBI gravity, as it has a complicated 
  spherically symmetric solution when the electromagnetic field is considered~\cite{12}. 
  \par
  In this paper, inspired by T. Padmanabhan~\cite{4,5,6}, we derive the first law of black hole 
  thermodynamics with the commonly accepted thermodynamics quantities from the Einstein's 
  equation. We also use this technique in EiBI gravity and get the known first law,
  the results show a more general formula of entropy, which also holds for 
  AdS Schwarzchild black hole and AdS R-N black hole. Motived by this, supposing a more 
  general static spherically symmetric metric, we get the same result in Einstein gravity.     
  \par
  This paper is organized as follows: In Sec.\uppercase\expandafter{\romannumeral2}, 
  there is a derivation of the thermodynamic identity from Einstein gravity, we also get a more 
  general formula of entropy.
  In Sec.\uppercase\expandafter{\romannumeral3}, we investigate the EiBI gravity by using the 
  thermodynamic route to the field equation, and get the formula of entropy in EiBI gravity. 
  Conclusions and discussion are given in Sec.\uppercase\expandafter{\romannumeral4}.          
  
  \section{BLACK HOLE THERMODYNAMIC IDENTITY FROM EINSTEIN'S EQUATION}
  
  The Einstein's equation can be derived from the thermodynamics~\cite{3}.
  On the other side, it is possible to interpret the Enistein's equation near the spherical 
  symmetric event horizon as the first law of thermodynamics which was proposed by 
  T. Padmanabhan~\cite{4,5,6}. However, the thermodynamic quantities might not be consistent 
  with the normal ones, especially the pressure and internal energy. Fortunately, through 
  restudying the field equation, we can also derive the first law of thermodynamics with the 
  commonly accepted thermodynamic quantities~\cite{24,30}.   
  
  Considering a static spherically symmetric spacetime 
  \begin{equation}
    \label{eq:metric-1}
    ds^2 = -f(r)dt^2 +\frac{1}{f(r)}dr^2+r^2(d\theta^2+sin^2\theta d\phi^2) \qquad,
  \end{equation}
  and the event horizon $r=r_{H}$ satisfying $f(r_{H})=0$, then one can get its thermodynamic 
  quantities
  \begin{equation}
    \label{eq:thermodynamic}
    T=\frac{\kappa}{2\pi}=\frac{f'(r_{H})}{4\pi},\qquad S=\pi r_{H}^2,\qquad V=\frac{4\pi}{3}
    r_{H}^3 .
  \end{equation}
  If we consider an AdS spacetime with a negative cosmological constant $\Lambda$, 
  there would be a pressure 
  $P=-\Lambda/8\pi$~\cite{24}. The mass of black hole was treated as enthalpy~\cite{30} and the first law 
  of black hole thermodynamics is 
  \begin{equation}
    \label{eq:first_law_H}
    dH=dM=TdS+VdP \qquad.
  \end{equation}
  Through the Legendre transformation, one can get the internal energy
  \begin{equation}
    \label{eq:first_law_U}
    dU=TdS-PdV \qquad.
  \end{equation}
  Especially, the internal energy for AdS Schwarzschild black hole and AdS R-N balck hole are
  \begin{equation}
    \label{eq:internal energy_S_R-N}
    U_{Schwarzschild}=\frac{r_{H}}{2},\qquad 
    U_{R-N}=\frac{r_{H}}{2}+\frac{Q^2}{2r_{H}}\qquad .
  \end{equation}
   
  \par
  The Einstein's equation with a cosmological constant is
  \begin{equation}
    \label{eq:Einstein's_equation}
    G_{\mu\nu}+\Lambda g_{\mu\nu}=8\pi T_{\mu\nu} \qquad.
  \end{equation}
  If the metric has the form of eq.\eqref{eq:metric-1}, in the case of $T_{\mu\nu}=0$, one can 
  obtain the $\theta\theta$ component of the equation
  \begin{equation}
    \label{eq:Einstein's_equation_2_component}
     -1+f(r)+rf'(r)=-\Lambda r^2 \qquad,
  \end{equation}
  Setting $r=r_{H}$, then multiplying 
  above equation by $dr_{H}$, we can rewrite eq.\eqref
  {eq:Einstein's_equation_2_component} as
  \begin{equation}
    \label{eq:Einstein's_equation_2_component_rH}
    \frac{f'(r_H)}{4\pi}d(\pi r_{H}^2)-d\left(\frac{r_H}{2}\right)=
    -\frac{\Lambda}{8\pi}d\left(\frac{4\pi r_{H}^3}{3}\right) \qquad.
  \end{equation}
  Noticing eq.\eqref{eq:thermodynamic}, above equation can be regarded as the first law of black 
  hole thermodynamics, since $U=r_{H}/2$ for the Schwarzschild solution. 
  \par
  For a charged AdS black hole, the metric also takes the form of eq.\eqref
  {eq:metric-1}. The energy-momentum tensor of electromagnetic field is 
  \begin{equation}
    \label{eq:energy_momentum_tensor_eletrovaccum}
     T_{\mu\nu}=\frac{1}{4\pi}(F_{\mu\sigma}F_{\nu}^{\sigma}-\frac{1}{4}g_{\mu\nu}F_{\sigma\rho}
     F^{\sigma\rho}) \qquad.
  \end{equation}
  Its none zero components are:
  \begin{eqnarray*}
  	\label{eq:energy_momentum_tensor_eletrovaccum_nonezero}
	T_{tt}&=&fE_{0}^2/8\pi,\qquad T_{rr}=-f^{-1}E_{0}^2/8\pi,\\
	T_{\theta\theta}&=&r^2E_{0}^2/8\pi,\qquad T_{\phi\phi}=r^2sin^2\theta E_{0}^2/8\pi,
  \end{eqnarray*}
  where $E_{0}=Q/r^2$, and $Q$ represents the charge of black hole. According to the Einstein's 
  equations, one can also get
  \begin{equation}
    \label{eq:Einstein's_equation_2_component_eletrovaccum}
     -1+f(r)+rf'(r)=-\Lambda r^2-r^2E_{0}^2 \qquad.
  \end{equation}
  By the same technique, and treating $Q$ as a constant, we get 
  \begin{equation}
    \label{eq:Einstein's_equation_2_component_eletrovaccum_rH}
     \frac{f'(r_H)}{4\pi}d(\pi r_{H}^2)-d\left(\frac{r_H}{2}+\frac{Q^2}
     {2r_{H}}\right)=Pd\left(\frac{4\pi r_{H}^3}{3}\right) \qquad,
  \end{equation}
  which can be also treated as the first law with the thermodynamic quantities in 
  eq.\eqref{eq:thermodynamic}, and one can 
  verify that $U=r_{H}/2+Q^2/2r_{H}$ for the AdS R-N black hole.
  \par
  Here we keep $Q$ as a constant, which means a chargeless particle falls into the AdS R-N black 
  hole, eq.\eqref{eq:Einstein's_equation_2_component_eletrovaccum_rH} is consistent with 
  the first law. 
  While a charged particle falls into the AdS R-N black hole, 
  the event horizon $r_{H}$ would arise due to changes of $dM$ and $dQ$, then eq.\eqref
  {eq:Einstein's_equation_2_component_eletrovaccum_rH} could be rewrite as~\cite{5}
  \begin{equation}
    \label{eq:Einstein's_equation_2_component_eletrovaccum_rH_Q}
     \frac{f'(r_H)}{4\pi}d(\pi r_{H}^2)-d\left(\frac{r_H}{2}+\frac{Q^2}
     {2r_{H}}\right)+\frac{Q}{r_{H}}dQ=Pd\left(\frac{4\pi r_{H}^3}{3}\right) 
     \qquad.
  \end{equation}
  Then it would adopt to the first law with the formulation
  \begin{equation}
    \label{eq:first_law_U_Q}
    dU=TdS-PdV+\Phi dQ \qquad.
  \end{equation}
  We should point out that the $T_{\mu\nu}$ contributes to the internal energy $U$.
  \par
  Thus the Einstein's equation can be interpreted as the first law of thermodynamic 
  near the event horizon. The technique was first proposed by T. Padmanabha, and some relevant 
  comments about the meaning of thermodynamic quantities for this result were 
  given~\cite{4,5,6,13}. 
  Since we used the different thermodynamics quantities in our derivation compared with 
  Padmanabha's work, there are some comments we would like to add. Firstly, this method can be 
  applied to single horizon and multiple horizons spacetime, but this is just a local 
  description of horizon thermodynamics which means the temperature, entropy and energy are 
  local quantities just for one horizon. In addition, This method still have some problem to 
  solve, when it was applied to cosmic horizon or de Sitter horizon, because the definition of 
  the temperatures for cosmic horizon is quite different. Secondly, it seems that there is a 
  manifest arbitrariness or freedom in the derivations.
  One could multiply the entire equation by an arbitrary function, then the expressions for 
  entropy, internal energy and volume would be another one. In fact, if we consider the 
  initial value, that $S=A/4$, $V=4\pi r_H^3/3$ for  Schwarzschild spacetime, this problem would 
  disappear. Finally, the perturbation of the static spacetime can be interpreted in terms of 
  the physical process of black hole evaporation or hawking radiation. One can also interpret 
  the relations by, say, dropping ‘test particles’ into the black hole.  
  
  Note that the structure of the equation itself allows us to read off the 
  expression for entropy. This technique has been used for 
  Gauss-Bonnet gravity and Lovelock gravity~\cite{5}, in which their entropy formulas are 
  the same with ~\cite{26,27}, respectively.
  we would like to consider a more general static spherical symmetric metric
  \begin{equation}
    \label{eq:metric-3}
    ds^2 = -\psi(r)^2 f(r)dt^2 +\frac{1}{f(r)}dr^2+r^2(d\theta^2+sin^2\theta d\phi^2) \qquad.
  \end{equation}
  One can calculate its nozero components of Ricci tensor
  \begin{align}
    \label{eq:Ricci_00}
    R_{tt}&=\frac{(\psi^2f)''f}{2}-\frac{(\psi^2f)'f}{4}\left(-\frac{f'}{f}+\frac{(\psi^2f)'}
    {\psi^2f}\right)+\frac{(\psi^2f)'f}{r} \qquad,\\
    \label{eq:Ricci_11}
    R_{rr}&=-\frac{(\psi^2f)''}{2\psi^2f}+\frac{(\psi^2f)'}{4\psi^2f}\left(-\frac{f'}{f}+
    \frac{(\psi^2f)'}
    {\psi^2f}\right)-\frac{f'}{rf} \qquad,\\
    \label{eq:Ricci_22}
    R_{\theta\theta}&=1-\frac{rf}{2}\left(\frac{f'}{f}+\frac{(\psi^2f)'}{\psi^2f}\right)-f
    \qquad,\\
    \label{eq:Ricci_33}
    R_{\phi\phi}&=sin^2\theta R_{\theta\theta} \qquad.
  \end{align}
  And the Einstein's equation with a cosmological constant can be written as 
  \begin{equation}
    \label{eq:Einstein's_equation_form2}
    R_{\mu\nu}-\Lambda g_{\mu\nu}=8\pi(T_{\mu\nu}-\frac{1}{2}Tg_{\mu\nu}) \qquad.
  \end{equation}
  \par
  For the Schwarzchild vacuum $T_{\mu\nu}=0$, one can get $\psi=C$, a constant. We can always 
  have $\psi =1$ by 
  choosing  the coordinate time $d\hat{t}=Cdt$ without changing the killing vector, 
  then the metric would back to 
  the spherically symmetric metric eq.\eqref{eq:metric-1}. For the charged 
  black hole, we have got the Reissner-Nordstrom metric, which also implies $\psi=1$,
  thus the first law could be obtained. The reason of this result might be that the 
  matter field leads to $\psi=1$. 
  \par 
  To get a general case, we just consider the $\theta\theta$ component of eq.\eqref
  {eq:Einstein's_equation_form2}, which can be expressed as
  \begin{equation}
    \label{eq:Einstein_equation_22}
    1-\frac{rf(r)}{2}\left(\frac{f'(r)}{f(r)}+\frac{(\psi^2f(r))'}{\psi^2f(r)}\right)-f(r)
    =\Lambda r^2+8\pi(T_{\theta\theta}-\frac{1}{2}Tr^2) \qquad.
  \end{equation}
  We assume the metric satisfies $f(r_{H})=0$ and $\psi(r_{H})\neq0$ on the event horizon.  
  By setting $r=r_{H}$ and considering the matter field contributes 
  to the internal energy $U$, then multiplying $dr_{H}$ one can get 
  \begin{equation}
    \label{eq:Einstein_equation_22_mod1}
    dU-\frac{r_{H}}{2}f'(r_{H})dr_{H}=-Pd\left(\frac{4\pi r_{H}^3}{3}\right) \qquad,
  \end{equation}
  where 
  \begin{equation}
    \label{eq:general_internal energy}
    dU=[\frac{1}{2}+4\pi(T_{\theta\theta}-\frac{1}{2}Tr_{H}^2)]dr_{H} \qquad.
  \end{equation}
  One can verify that for the AdS Schwarzchild black hole, it reduces to $U=r_{H}/2$ and for the 
  AdS R-N black hole, it gives $U=r_{H}/2+Q^2/2r_{H}$. So the internal energy expression is just 
  a generalization.
  \par
  The Hawking temperature on the event horizon becomes
  \begin{equation}
    \label{eq:surface_gravity}
     T=\frac{\kappa}{2\pi}=\frac{\psi(r_{H})f'(r_{H})}{4\pi} \qquad.
  \end{equation}   
  Now rewrite eq.\eqref{eq:Einstein_equation_22_mod1} as 
  \begin{equation}
    \label{eq:Einstein_equation_22_mod2}
    dU-T\left(\frac{2\pi r_{H}}{\psi(r_{H})}dr_{H}\right)=-PdV \qquad.
  \end{equation}
  One would 
  find the entropy has to satisfy
  \begin{equation}
    \label{eq:Entropy_mod}
    dS=\frac{2\pi r_{H}}{\psi(r_{H})}dr_{H} \qquad,
  \end{equation} 
  or
  \begin{equation}
    \label{eq:Entropy_mod_int}
    S=\int{\frac{2\pi r_{H}}{\psi(r_{H})}}dr_{H} \qquad.
  \end{equation} 
  Thus, we generalize the corrected entropy formula to Einstein gravity for a static 
  spherically symmetric metric eq.\eqref{eq:metric-3}.
  Once $\psi=1$, it is obvious that $S=\pi r_{H}^2=A/4$, which is the well-known 
  Bekenstein-Hawking black hole entropy. For the Schwarzschild black hole and R-N black hole, 
  which all have $\psi=1$, $S=A/4$. 
  \par
  In fact, Matt Visser has proposed that the entropy of `dirty' black holes might not equal 
  quarter of area of event horizon~\cite{28,29}. Generically, a `dirty' black hole is a black 
  hole with various classical matter fields, higher curvature terms in the gravity Lagrangian 
  or some other versions of quantum hair. For the Einstein gravity, a more 
  general metric eq.\eqref{eq:metric-3} could be caused by some special matter fields like 
  electromagnetism with dilaton fields~\cite{7}. So the entropy also should be corrected in 
  Einstein gravity. In the next section, we would like to consider the EiBI gravity~\cite{8}, 
  its Lagrangian includes the higher curvature terms in eq.\eqref{eq:eibi_action}, which could 
  be interpreted as the self-gravity, we will show our discussion of this method in the 
  Eddington-inspired Born-Infeld gravity and get its thermodynamic quantities.    
     
  \section{THE ENTROPY IN EDDINGTON-INSPIRED BORN-INFELD GRAVITY}
  
  The Eddington-inspired Born-Infeld theory of gravity is based on the Palatini formulation 
  which treats the metric and connection as independent fields~\cite{8}. Its action can be 
  written as 
  \begin{equation}
    \label{eq:eibi_action}
    S=\frac{1}{8\pi\kappa}\int{d^4x[\sqrt{|g_{\mu\nu}+\kappa R_{\mu\nu}(\Gamma)|}-
    \lambda\sqrt{g}]}+S_{M}(g,\Gamma,\Psi) \qquad,
  \end{equation}
  where $g_{\mu\nu}$ is the metric of spacetime and its determinant is $g$, $R_{\mu\nu}$ is 
  the symmetric Ricci tensor related to $\Gamma$, the dimensionless parameter $\lambda=1+\kappa
  \Lambda$, and the parameter $\kappa$ has the inverse dimension of 
  cosmological constant $\Lambda$.
  \par
  By varying the action with respect to $g_{\mu\nu}$ and $\Gamma$, one obtains the equation of
  motion 
  \begin{equation}
    \label{eq:eibi_equation_1}
    q_{\mu\nu}=g_{\mu\nu}+\kappa R_{\mu\nu} \qquad,
  \end{equation}   
  \begin{equation}
    \label{eq:eibi_equation_2}
    \sqrt{|q|}q^{\mu\nu}=\lambda\sqrt{|g|}g^{\mu\nu}-8\pi\kappa\sqrt{|g|}T^{\mu\nu} \qquad,
  \end{equation}
  where $q_{\mu\nu}$ is the auxiliary metric compatible to the connection with 
  \begin{equation}
    \label{eq:eibi_connection}
    \Gamma_{\mu\nu}^{\lambda}=\frac{1}{2}q^{\lambda\sigma}(q_{\mu\sigma,\nu}+q_{\nu\sigma,\mu}-
    q_{\mu\nu,\sigma}) \qquad.
  \end{equation}
  By combining eq.\eqref{eq:eibi_equation_1} and eq.\eqref{eq:eibi_equation_2}, then expanding 
  the field equations to 2nd order of $\kappa$~\cite{8} 
  \begin{equation}
    \label{eq:eibi_equation_expanding}
    R_{\mu\nu} \simeq \Lambda g_{\mu\nu}+8\pi(T_{\mu\nu}-\frac{1}{2}Tg_{\mu\nu})+
    8\pi\kappa[S_{\mu\nu}-\frac{1}{4}Sg_{\mu\nu}]\qquad,
  \end{equation}
  where $S_{\mu\nu}=T_{\mu}^{\alpha}T_{\alpha\nu}-\frac{1}{2}TT_{\mu\nu}$, one can find that 	
  the equation is the 1st order corrections to Einstein's equation. On the other hand, EiBI 
  gravity can be interpreted as a correction of the matter term compared with Einstein gravity. 
  Even the EiBI gravity is fully equivalent to the Einstein gravity in vacuum.
  \par
  Let's consider the thermodynamics from the field equation in this gravity model. 
  Generally, a static spherically symmetric metric $g_{\mu\nu}$ could be
  \begin{equation}
    \label{eq:metric-2}
    ds_{g}^2 = -\psi^2(r) f(r)dt^2 +\frac{1}{f(r)}dr^2+r^2(d\theta^2+sin^2\theta d\phi^2) \qquad,
  \end{equation}
  and the auxiliary metric $q_{\mu\nu}$ is assumed as~\cite{12} 
  \begin{equation}
    \label{eq:metric-4}
    ds_{q}^2 = -G^2(r) F(r)dt^2 +\frac{1}{F(r)}dr^2+H^2(r)(d\theta^2+sin^2\theta d\phi^2) \qquad.
  \end{equation}
  The Ricci tensor was calculated as below 
  \begin{align}
  \label{eq:Ricci_00_q}
  	R_{tt}&=2\frac{GG'H'F^2}{H}+\frac{G^2FF'H'}{H}+\frac{3}{2}GG'FF'+GG''F^2+ 
  	\frac{1}{2}G^2FF'' \qquad,\\
  \label{eq:Ricci_11_q}
     R_{rr}&=-2\frac{H''}{H}-\frac{F'H'}{FH}-\frac{3}{2}\frac{G'F'}{GF}-\frac{G''}{G}-
    \frac{F''}{2F} \qquad,\\
  \label{eq:Ricci_22_q}
    R_{\theta\theta}&=1-HH'F'-\frac{G'}{G}HH'F-H'^2F-HH''F  \qquad,\\
  \label{eq:Ricci_33_q}
    R_{\phi\phi}&=sin^2\theta R_{\theta\theta}\qquad,
  \end{align}
  where the $X'=\partial X /\partial r$, which we will use in the rest of this paper.
  \par  
  Without the matter fields, the eq.\eqref{eq:eibi_equation_2} reduces to  
  \begin{align}
  	\label{eq:eibi_equation_2_components_vacuum_tt}
  	\frac{H^2}{GF}&=\frac{\lambda r^2}{\psi f}  \qquad,\\
  	\label{eq:eibi_equation_2_components_vacuum_rr}
  	GH^2 F&=\lambda r^2\psi f  \qquad,\\
  	\label{eq:eibi_equation_2_components_vacuum_cc}
  	G&=\lambda \psi  \qquad,
  \end{align}
  thus one can obtian
  \begin{equation}
  G=\lambda\psi,\qquad F=\lambda^{-1}f,\qquad H^2=\lambda r^2.\qquad
  \end{equation}
  Plugging these into eq.\eqref{eq:eibi_equation_1},
  then the $\theta\theta$ component is
  \begin{equation}
    \label{eq:eibi_equation_1_2_component}
     1-rf'(r)-(\frac{r\psi'}{\psi}+1)f(r)= \frac{1}{\kappa}(\lambda-1)r^2 \qquad.
  \end{equation}
  Near the event horizon $r=r_{H}$ ($f(r_{H})=0$), we have 
  \begin{equation}
    \label{eq:eibi_equation_1_2_component_rH}
     d\left(\frac{r_{H}}{2}\right)-\frac{f'(r_{H})}{4\pi}d(\pi r_{H}^2)=-Pd\left(
     \frac{4\pi r_{H}^3}{3}\right)\qquad.
  \end{equation}
  As the EiBI gravity is fully equivalent to the Einstein gravity in vacuum~\cite{8}, 
  the above equation could imply the first law. In fact, the black hole solution to EiBI gravity 
  with no source is the same as Schwarzchild-de Sitter metric, which illustrates 
  $\psi=1$ and $f=1-2m/r+\Lambda r^2/3$~\cite{8}. The thermodynamic quantities in 
  eq.\eqref{eq:eibi_equation_1_2_component_rH} also hold in EiBI 
  gravity. Thus the first law can also be 
  got from the 
  EiBI gravity. Next, we would consider the EiBi gravity with electromagnetic field.    
  \par
  The energy-momentum tensor of electromagnetic field could be  
  \begin{eqnarray*}
  	\label{eq:energy_momentum_tensor_eletrovaccum_nonezero_mod}
	T^{tt}&=&(\psi^2f)^{-1}E_{0}^2/8\pi,\qquad T^{rr}=-fE_{0}^2/8\pi,\\
	T^{\theta\theta}&=&r^{-2}E_{0}^2/8\pi,\qquad T^{\phi\phi}=r^{-2}sin^{-2}\theta E_{0}^2/8\pi,
  \end{eqnarray*}
  where $E_{0}=Q/r^2$ and the $Q$ represents the charge of black hole. Then the eq.\eqref
  {eq:eibi_equation_2} becomes 
  \begin{align}
  	\label{eq:eibi_equation_2_components_vacuum_tt}
  	\frac{H^2}{GF}&=(\lambda+\kappa E_{0}^2)\frac{r^2}{\psi f}  \qquad,\\
  	\label{eq:eibi_equation_2_components_vacuum_rr}
  	GH^2 F&=(\lambda+\kappa E_{0}^2)r^2\psi f  \qquad,\\
  	\label{eq:eibi_equation_2_components_vacuum_cc}
  	G&=(\lambda-\kappa E_{0}^2) \psi  \qquad,
  \end{align}
  and one can get 
  \begin{equation}
  	G=(\lambda-\kappa E_{0}^2)\psi,\qquad F=(\lambda-\kappa E_{0}^2)^{-1}f,\qquad 
  	H^2=(\lambda+\kappa E_{0}^2)r^2.
  \end{equation}
  The $\theta\theta$ component of eq.\eqref{eq:eibi_equation_1} is written as 
  \begin{equation}
  \label{eq:eibi_equation_1_2_component_eletrovaccum}
  	1-\frac{\lambda+\kappa E_{0}^2+\kappa r E_{0}E_{0}'}{\lambda-\kappa E_{0}^2}\cdot rf'
  	-\frac{Y}{\lambda-\kappa E_{0}^2}f= \frac{1}{\kappa}(\lambda-1)r^2+E_{0}^2r^2
  	\qquad,
  \end{equation}
  where
  \begin{equation}
    \label{eq:eibi_equation_1_2_component_eletrovaccum_Y}
  	Y=2\kappa rE_{0}E_{0}'+\frac{G'}{G}HH'-H'^2
  	\qquad.
  \end{equation}
  If we assume that the event horizon satisfies $f(r_{H})=0$ and $\psi(r_{H})\neq 0$, 
  then set $r=r_{H}$ in eq.\eqref{eq:eibi_equation_1_2_component_eletrovaccum} and multiply it 
  by $dr_{H}$, noting $E_{0}'=-2E_{0}/r$, it gives
  \begin{equation}
  \label{eq:eibi_equation_1_2_component_eletrovaccum_rH}
  d\left(\frac{r_{H}}{2}+\frac{Q^2}{2r_{H}}\right)-\frac{r_{H}}{2}f'(r_{H})dr_{H}=
  -Pd\left(\frac{4\pi r_{H}^3}{3}\right) \qquad.
  \end{equation}
  This equation should be the first law since it can go back to eq.\eqref
  {eq:eibi_equation_1_2_component_rH} when $Q=0$. And one can also confirm this by noticing that 
  $dU=d(r_{H}/2+Q^2/2r_{H})$ and $dV=d(4\pi r_{H}^3/3) $. Moreover, it gives the same result in 
  Einstein gravity. So we should 
  identify that 
  \begin{equation}
    \label{eq:eibi_TdS_term}
    TdS=\frac{r_{H}}{2}f'(r_{H})dr_{H} \qquad.
  \end{equation}
  \par
  However, as the metric takes the form 
  of eq.\eqref{eq:metric-2}, the surface gravity 
  could be~\cite{7}:
  \begin{equation}
    \label{eq:surface_gravity}
     \kappa=\lim_{r\to r_{H}} \frac{1}{2}\frac{\partial_{r}g_{tt}}{\sqrt{g_{tt}g_{rr}}}=
     \frac{\psi(r_{H})f'(r_{H})}{2} \qquad.
  \end{equation}
  One can obtain the temperature on the event horizon
  \begin{equation}
    \label{eq:eibi_temperature_eletrovaccum}
  	T=\frac{\psi(r_{H})f'(r_{H})}{4\pi} \qquad.
  \end{equation} 
  Then eq.\eqref{eq:eibi_TdS_term} would imply that
  \begin{equation}
	\label{eq:eibi_temperature_entropy}
     dS=\left(\frac{2\pi r_{H}}{\psi(r_{H})}\right)dr_{H} \qquad,
  \end{equation}    
  or
  \begin{equation}
	\label{eq:eibi_temperature_entropy_int}
     S=\int{\frac{2\pi r_{H}}{\psi(r_{H})}dr_{H}} \qquad.
  \end{equation} 
  Obviously, when $\psi=1$, one can get $S=\pi r_{H}^2$.
  \par
  Thus, we get the entropy formula in EiBI gravity. Surprisingly, we find it also holds 
  in Einstein gravity once metric takes the form of eq.\eqref{eq:metric-2}. Therefore, we can 
  also get the first law for a more general static spherically symmetric 
  metric in EiBI gravity. Moreover, one can find that eq.\eqref
  {eq:Einstein's_equation_2_component_eletrovaccum_rH} is the same as eq.\eqref
  {eq:eibi_equation_1_2_component_eletrovaccum_rH}.
  It implies that the Einstein gravity and EiBI gravity 
  might be equivalent on the event horizon from the view of black hole thermodynamics.   
  \par
  In fact, the black hole solution with electromagenatic field in EiBI gravity has been found, 
  while $f(r_{H})=0$ and $\psi(r_{H})\neq 0$~\cite
  {8,12}. It is given as below
  \begin{equation}
  \label{eq:eibi_psi}
  	\psi(r)=\frac{\sqrt{\lambda}r^2}{\sqrt{\lambda r^4+\kappa Q^2}} \qquad.
  \end{equation}   
  With this result we can get the corrected entropy
  \begin{align}
  	S&= \int{\frac{2\pi r_{H}}{\psi(r_{H})}}dr_{H}=\pi\int{\frac{1}{r_{H}^2}\sqrt{r_{H}^4+
  	\frac{\kappa}{\lambda}Q^2}} dr_{H}^2 \\
  	\label{eq:eibi_temperature_entropy_int_electrovacuum}
  	 &=\pi\sqrt{r_{H}^4+\frac{\kappa}{\lambda}Q^2}-\pi\sqrt{\frac{\kappa}{\lambda}}|Q|
  	\cdot
  	ln\left(\sqrt{\frac{\kappa}{\lambda}}\frac{|Q|}{r_{H}^2}+\sqrt{1+\frac{\kappa}
  	{\lambda}\frac{Q^2}{r_{H}^4}}\right) \qquad.
  \end{align}   
  A logarithmic term occurs in this formula as a corrected entropy.
  When $Q=0$, one gets $S=\pi r_{H}^2$, which is consistent with the vacuum case, 
  and so is the same as Einstein gravity. When $\kappa\to 0$, EiBI gravity would reduce to 
  the Einstein gravity, and entropy becomes the Bekenstein-Hawking one. 
  
  \section{CONCLUSION And DISCUSSION}
  
  In this paper, we restudied T. Padmanabhan's work that it is possible to write Einstein's 
  equation for a spherically symmetric spacetime in the form of the first law of 
  thermodynamics~\cite{4,5,6,13}, but the thermodynamic quantities might not be consistent 
  with the normal ones, especially the pressure and internal energy. By using this technique, 
  we reproduced the first law with the commonly accepted thermodynamic quantities in the AdS 
  spacetime, and this technique provide an effective approach to read off the thermodynamic 
  quantities. Next, we investigated a more general static spherically symmetric metric taking
  the form eq.\eqref{eq:metric-3} in Einstein gravity. It is surprised to find that the entropy 
  might have a correction in Einstein gravity.
  \par
  Since it provided a convenient approach to study the black hole thermodynamics just from the 
  field equation, we 
  investigated the black hole thermodynamic in EiBI gravity. We found that there is nothing 
  different from Einstein gravity in vacuum, but entropy could be different from the R-N 
  black hole when the electromagnetic field was considered. 
  The corrected entropy from EiBI gravity should be eq.\eqref
  {eq:eibi_temperature_entropy_int}, which can reduce to the Bekenstein-Hawking entropy when 
  $\psi=1$, and it is also the same as Einstein gravity when $Q=0$ without the matter 
  field~\cite{8}. Thus, the corrected entropy in Einstein gravity could be generalized to EiBI 
  gravity. 
  \par
  Moreover, as the Einstein gravity and EiBI gravity hold the same result, 
  we remarked that these two theories of gravity could be equivalent
  on the event horizon from the view of thermodynamics.        
  \par
  At last, as an example, a specific corrected entropy of the charged black hole in EiBI gravity 
  was given. The entropy form would lead something different for the black hole thermodynamics 
  in EiBI gravity, like the phase transition and critical phenomenon. These would be left for 
  our further research. 
  
  \section*{Conflict of Interests}
  The authors declare that there is no conflict of interests regarding the publication of this 
  paper.
  
  \section*{ACKNOWLEDGMENTS}
  We would like to thank the National Natural Science Foundation of
  China~(Grant No.11571342) for supporting us on this work.

  \bibliographystyle{unsrt}
  \bibliography{reference}

\end{document}